\documentclass[prb,aps,floats,amssymb,showkeys,showpacs,preprint,
 bibnotes, tightenlines]{revtex4}
\usepackage{longtable}
\usepackage{graphicx}
\usepackage{graphics}
\usepackage{epsfig,bm}
\usepackage{rotating}
\begin{document}

\title{The high-temperature expansions of the
 higher susceptibilities\\ for the Ising model in general dimension $d$}

\author{P. Butera}

\email{paolo.butera@mib.infn.it}

\affiliation
{Dipartimento di Fisica Universita' di Milano-Bicocca\\
and\\
Istituto Nazionale di Fisica Nucleare \\
Sezione di Milano-Bicocca\\
 3 Piazza della Scienza, 20126 Milano, Italy}

\author{M. Pernici} 

\email{mario.pernici@mi.infn.it}

\affiliation
{Istituto Nazionale di Fisica Nucleare \\
Sezione di Milano\\
 16 Via Celoria, 20133 Milano, Italy}

\date{\today}

\date{\today}
\begin{abstract}
 The high-temperature expansion coefficients of the ordinary and the
 higher susceptibilities of the spin-1/2 nearest-neighbor Ising model
 are calculated exactly up to the 20th order for a general
 $d$-dimensional (hyper)-simple-cubical lattice.  These series are
 analyzed to study the dependence of critical parameters on the
 lattice dimensionality. Using the general $d$ expression of the
 ordinary susceptibility, we have more than doubled the length of the
 existing  series expansion of the critical temperature
 in powers of $1/d$.
\end{abstract}
\maketitle

\section{Introduction}

 We have derived high-temperature(HT) expansions of the ordinary and
the higher susceptibilities (see the definitions in Section II)
 of the spin-1/2 Ising model exactly up to
the 20th order for the general $d$-dimensional (hyper)-simple-cubical(hsc)
lattices.

 These expressions are obtained by interpolation of the HT series
 expansion coefficients of the susceptibilities over the integer
 values of the lattice dimensionality, and not by analytic
 continuation. Thus they have no obvious uniqueness properties when
 $d$ is allowed to take noninteger values. {\it A priori} a
 different dependence of physical quantities on the dimensionality
 might result from different possible interpolations, such as that
 obtained
%by the $\epsilon=4-d$ expansion\cite{zinnbook,zinleg} of the
% renormalization group theory or by the fixed-dimension
% renormalization group\cite{holo} or 
by formulating the Ising model on
 a fractal lattice\cite{gef,bonnier}, whose Hausdorff dimension can be
 varied continuously, or also in other ways\cite{novot}.  An analogous
 situation is known to occur for the $n$-vector model, whose HT series
 coefficients can be interpolated\cite{bcoenne} by rational functions
 of $n$.  Of course, non-integer values of $d$ (or similarly of $n$)
 might in some cases have no physical meaning\cite{baker}.

   The expansions\cite{fishergaunt, gerberfisher,gaunt} of the
physical quantities in powers of $1/d$, i.e.  around the
{\it mean-field} (MF) approximation (or in powers\cite{zinnbook} of $1/n$,
i.e. the expansions around the spherical model limit), are related
with these analytic representations in terms of $d$ (or of $n$).

Our results not only provide reference data in a compact form for the
 higher susceptibilities, which are generally difficult to compute by
 methods different from series expansions, but also make a variety of
 other investigations possible. For example, in
 discussing\cite{luijten,lundow,mer5,mer6,ak6,ak7,ak8} how the
 finite-size-scaling behavior\cite{chen} changes for systems above the
 upper critical dimensionality, accurate estimates of the critical
 parameters are needed as benchmarks.  Our data can also help to
 assess the accuracy of estimates of physical parameters obtained from
 approximations of a different nature, such as the the $\epsilon=4-d$
 expansion\cite{zinnbook,zinleg} of the renormalization group theory,
 the fixed-dimension renormalization group\cite{holo}, the
 $1/d$-expansion\cite{fishergaunt,gaunt}, the MonteCarlo(MC)
 simulations, etc..  It is appropriate at this point to observe that
 the MC simulations become increasingly time- and memory-demanding as
 the lattice dimensionality $d$ grows, whereas the non-analytic
 corrections to scaling in the asymptotic critical behavior of physical
 quantities, that usually make the HT series analyses a delicate
 matter and are the main source of their uncertainties, become simpler
 and smaller\cite{guttm} with increasing $d$.  Thus even moderately
 long HT expansions can lead to very accurate estimates already for not
 very large $d$.

It is also worth  mentioning that extremely long, although approximate
 HT expansions of the ordinary susceptibility have been recently
 generated\cite{berche} by a MC method for the hsc lattices of
 dimensionalities $d=5,..,8$ and used to test the accuracy of the
 ``extended scaling'' ideas\cite{camp1,camp2,camp3} above the upper
 critical dimension.  The results of this investigation can now be
 compared with those from the analysis of our far shorter, but exact
 expansions.

 The expressions presented here are obtained from recently
 derived\cite{bp1,bp2,bp3} HT and low-field series expansions of the
 magnetization in presence of an external magnetic field for the
  spin-1/2 Ising model with nearest-neighbor
 interactions. Actually, we have produced a wider ranging set of data
 including as well other spin-systems in the Ising model universality
 class, such as the general spin-$s$ Ising model and the lattice
 scalar-field theories with polynomial self-interaction and thus, also
 for these models we are able to write exact expressions valid for
 general $d$-dimensional hsc lattices.

 Our derivation of the HT and low-field series, that significantly extend the
 longest known results, even in zero
 field\cite{bcoenne,bcisiesse,bcisiesse2,bcfi4,bcgrenold,bcgren}, have
 been performed for the lattice dimensionalities $d=1,2,...10$. The
 expansions are carried to the 24th order in the case of the
 (hyper)-body-centered-cubical lattices, but they are slightly shorter
 in the case of the hsc
 lattices: we have obtained the 24th order for $d<5$, the 22nd for
 $d=5$, the 21st for $d=6$, the 20th for $7 \leq d \leq 10$.

The layout of the paper is the following: in the second Section, we
set our notations and tabulate 
 the expansion coefficients of the ordinary susceptibility as
closed-form polynomials in the coordination number $q=2d$.   The
corresponding data for the higher susceptibilities are reported
 in the appendix. The printed version of this paper\cite{bpd}
 contains no appendix and the data for the higher susceptibilities
 are retrievable 
as Supplemental Material at the URL: 
 http://link.aps.org/supplemental/10.1103/PhysRevE.86.011139.
  In the third Section, we discuss the results of some analyses
of these series. The last Section contains our conclusions.

\section{ The Ising model in general dimension}
 The partition function of a spin-1/2 Ising system with
 nearest-neighbor interactions, in the presence of an external
 magnetic field $H$, on a finite $d$-dimensional 
lattice of $N$ sites can be written
 as
\begin{equation}
Z_N(H,T;d)=\sum_{conf} 
{\rm exp}\Big [ J/k_BT \sum_{<ij>} s_is_j +H/k_BT \sum_i s_i \Big ]. 
\label{Isipf}
\end{equation}

Here $s_i= \pm 1$ denotes an Ising spin variable associated to the
 lattice site $i$.  The first sum extends to all configurations of the
 spins, the second to all distinct pairs $<ij>$ of nearest-neighbor
 spins and the third to all spins. We shall set $K= J/k_BT$, with $T$ the
 temperature, $J$ the exchange coupling, $k_B$ the Boltzmann
 constant, and $h=H/k_BT$ the reduced magnetic field.  In terms of the
 variable $v={\rm tanh}K$, the HT expansion coefficients are simple
 integers and so this variable is more convenient  for the data
 tabulation.

In the thermodynamic limit $N \rightarrow \infty$, the specific free-energy
${\cal F}(h,K;d)$  is defined by
\begin{equation}
-\frac {K}{J}  {\cal F}(h,K;d) =\lim_{N \rightarrow \infty} \frac{1}{N} 
{\rm ln} Z_N(h,K;d)
\label{spfe}
\end{equation} 
The specific magnetization ${\cal M}(h,K;d)$  is defined by
\begin{equation}
 {\cal M}(h,K;d) =-\frac {K} {J}  \frac{\partial {\cal F}(h,K;d) } {\partial h}
\label{spmag}
\end{equation} 
Our HT calculation of the field-dependent magnetization, has yielded
 significant extensions of the existing HT expansions in zero field for 
the $2p$-spin connected-correlation-functions at zero wave-number and
zero field $\chi_{2p}(K;d)$ (usually 
called {\it ``higher susceptibilities''}). 
These quantities are 
defined by the successive field-derivatives of the specific magnetization
\begin{equation}
 \chi_{2p}(K;d)=(\partial^{2p-1} {\cal M}(h,K;d)/\partial h^{2p-1})_{h=0}
=\sum_{s_2,s_3,...,s_{2p}}<s_1 s_2...s_{2p}>_c.
\label{ncorr}
\end{equation}
 at zero field.

  The even field-derivatives of the magnetization vanish at zero
 field in the symmetric HT phase, while all  derivatives are nontrivial
 in the broken-symmetry low-temperature phase.

 The HT expansion coefficients of the susceptibilities at a given
order $l$ in $K$ are usually computed as sums of contributions classified in
terms of graphs having $l$ edges. To each graph we associate a weight
depending on the symmetry of the graph and on its free-multiplicity,
i.e. the number of distinct ways (per lattice site) in which the graph
can be embedded in the lattice, associating each vertex to a site and
each line to a nearest-neighbor bond\cite{wortis}.  Only the latter
quantity, technically denoted as {\it ``free-lattice-embedding number''}
within the linked-cluster HT expansion method, depends on the lattice
dimensionality.  An analysis of these numbers for the various classes
of graphs, like that performed in
Refs.[\onlinecite{fishergaunt,gerberfisher}] leads to the conclusion
that, at any given expansion order $l$, the HT series coefficients of
the ordinary and the higher susceptibilities can be written as simple
polynomials in the lattice dimensionality $d$ of degree at most $l$.
It is then clear that a straightforward prescription to represent the
HT series coefficients of a susceptibility as polynomials in $d$ up to
the order $K^{l_{max}}$, consists in repeating the computation of the
HT series for lattices of dimensionalities $d=0,1,...,l_{max}$ and
then in interpolating each series coefficient with respect to $d$.
Unfortunately, this straightforward strategy works only for relatively
small orders of expansion, since, in the case of 
 the hsc lattice the combinatorial complexity of the
computation of the graph-embedding numbers grows large exponentially
with the dimensionality.

Here we take advantage of a well known result that is helpful in
mitigating this difficulty.  It was shown long ago\cite{domb} that the
HT expansions of the successive derivatives of the magnetic field with
respect to the magnetization $ \partial^{2p+1} h/\partial {\cal
M}^{2p+1}$ at zero magnetization, for $p=0,1...$ contain only star
graphs, i.e. connected graphs having no articulation vertex. This property can
be very simply understood observing that in field-theoretic language
these derivatives define the connected amputated
one-particle-irreducible correlations at zero wave-number.

This property was sometimes used to restrict the number of graphs
contributing to the HT expansion of higher susceptibilities. What is
interesting for our aims is the fact that the lattice-embedding number
of a star graph with $l$ edges is a polynomial in $d$ of degree
$[l/2]$ at most. Here $[l/2]$ denotes the integer part of  $l/2$. 
The higher susceptibilities are simply related to the
quantities $ \partial^{2p+1} h/\partial {\cal M}^{2p+1}$:

\begin{equation}
\frac {\partial h} {\partial {\cal M}} (K;d)=\frac{1} {\chi_2(K;d)}
\label{chi2}
\end{equation}

\begin{equation}
\frac {\partial^3 h} {\partial {\cal M}^3}(K;d)
=\frac{\chi_4(K;d)}{\chi_2(K;d)^4}
\label{chi4}
\end{equation}

\begin{equation}
\frac{\partial^5 h} {\partial {\cal M}^5}(K;d)
=\frac{\chi_6(K;d)}{\chi_2(K;d)^6} - 10 \frac{\chi_4(K;d)^2}{\chi_2(K;d)^7}
\label{chi6}
\end{equation}

\begin{eqnarray}
\frac{\partial^7 h} {\partial {\cal M}^7}(K;d)
= \frac{\chi_8(K;d)}{\chi_2(K;d)^8} 
- 56\frac{\chi_4(K;d)\chi_6(K;d)}{\chi_2(K;d)^9}
+ 280\frac{\chi_4(K;d)^3}{\chi_2(K;d)^{10}} 
\label{chi8}
\end{eqnarray}

\begin{eqnarray}
\nonumber \frac{\partial^9 h} {\partial {\cal M}^9}(K;d)= 
 \frac{\chi_{10}(K;d)}{\chi_2(K;d)^{10}} 
-120\frac {\chi_4(K;d) \chi_8(K;d)}{\chi_2(K;d)^{11}}  
- 126\frac{\chi_6(K;d)^2}{\chi_2(K;d)^{11}} \\
+ 4620\frac{\chi_4(K;d)^2 \chi_6(K;d)}{\chi_2(K;d)^{12}}
-15400\frac{\chi_4(K;d)^4}{\chi_2(K;d)^{13}} 
\label{chi10}
\end{eqnarray}

\begin{eqnarray}
\nonumber \frac{\partial^{11} h} {\partial {\cal M}^{11}}(K;d)
=  \frac{\chi_{12}(K;d)}{\chi_2(K;d)^{12}}
- 220\frac{\chi_4(K;d)\chi_{10}(K;d)}{\chi_2(K;d)^{13}}
-  792\frac{\chi_6(K;d)\chi_8(K;d)}{\chi_2(K;d)^{13}}\\
\nonumber  + 17160\frac{\chi_4(K;d)^2\chi_8(K;d)}{\chi_2(K;d)^{14}} 
+ 36036\frac{\chi_4(K;d)\chi_6(K;d)^2}{\chi_2(K;d)^{14}} 
 - 560560\frac{\chi_4(K;d)^3\chi_6(K;d)}{\chi_2(K;d)^{15}} \\
 + 1401400\frac{\chi_4(K;d)^5}{\chi_2(K;d)^{16}}
\label{chi12}
\end{eqnarray}

 and so on.

 Then it transpires that it is sufficient to interpolate the HT series
expansion coefficients of the combinations of (higher) susceptibilities
on the rhs of eqs.(\ref{chi2}),...,(\ref{chi12}) etc. only over the
dimensionalities $1\le d \le [l/2]$, to obtain the representations of
these quantities for general $d$ through the $l$-th order in $K$. Finally, from
these results the representations of the single higher
susceptibilities can be easily recovered.  This simple remark leads to
a decisive reduction of the combinatorial complexity of the necessary
calculations. In our case, only the knowledge of the HT expansions 
of the (higher) susceptibilities for
all hsc lattices with $ d \le 10$ up to the 20th order is sufficient
to obtain the expression of these susceptibilities for general $d$ up to
the same order. The fact that the coefficient of order $K^l$ in
$\chi_{2p}(K;d)$ must be a polynomial in $d$ of degree $l$, while the
corresponding coefficient at the same order in the expansion of $
\partial^{2p-1} h/\partial {\cal M}^{2p-1}$ is a polynomial of degree
[$l$/2], provides a simple consistency check of our computations.

 A brief technical comment on a detail of the calculation is
appropriate at this point, since most of the computing time goes into
counting the number of lattice embeddings of each graph for relatively
large lattice dimensionality.  The first step of the computation
consists in ordering appropriately the vertices of the graph and in
placing the first of them at the lattice origin. The second step
consists in counting the possible positions, of coordinates
$(x_1,..,x_d)$, for the second vertex.  It is crucial to optimize this
step by using the hypercubical symmetry to restrict the possible
positions of the second vertex to the fundamental region $x_1 \geq x_2
\geq .. \geq x_d \geq 0$.  Separating each $\geq$ case into a $>$ and
an $=$, one gets $2^d$ cases.  The case with all $>$ corresponds to
the inside of the fundamental region, whose sites are representatives
of a group orbit of length $d! 2^d$.  A small program precomputes the
length of the orbits for the each of the $2^d$ cases, which is then
used in the embedding program.  After fixing the first two points of
the embedding, the possible positions of the remaining vertices are
restricted to relatively few sites by the constraints given by the
distances from the first two points.

 The timings for computing the HT expansion of the $d$-dimensional
 Ising model at order $l$ have been roughly $O(5.5^l 2.5^d)$.  In
 particular, the $10$-dimensional Ising model at order $20$ took $42$
 days of single-core time on a quad-core computer with a CPU-clock 
 frequency of $2.8GHz$.

\section{The HT expansions} 
 
 Expressions like those obtained here have already appeared at lower
orders in the literature for the spin-1/2  Ising
model\cite{fishergaunt,bakerd,gofm, bend} with nearest-neighbor interaction.
  In the case of the
ordinary susceptibility $\chi_2(K;d)$, they reached at most\cite{gofm}
the 15th order, while for $\chi_4(K;d)$ and $\chi_6(K;d)$ they were
computed\cite{bend} only up to the 11th order.

For brevity, we shall report our results in the Table \ref{tab1} only
   for $\chi_2(K;d)$ in the case of the spin-1/2 Ising model.
 The corresponding data for the higher
   susceptibilities $\chi_4(K;d)$,...$\chi_{16}(K;d)$ are  reported
   in the Appendix.

   We shall tabulate here also  the coefficients of order $< 16$,
   although they reproduce those already listed in
   Ref.[\onlinecite{gofm}]. A simplification of the expressions of the
   coefficients is obtained by using the variable $q=2d$ rather than
   the variable $d$ used in Ref.[\onlinecite{gofm}]. As already pointed out,
   we can produce analogous formulas also for the other models in the
   Ising universality class for which we have computed the HT
   expansions of the magnetization, but they will not be presented
   here. 
\begin{center}
%\begin{longtable}{|l|}
\begin{longtable}{l}
\caption {The coefficients $c^{(2)}_n(d)$ of the HT series expansion in
powers of $v={\rm tanh}K$ for the ordinary susceptibility
$\chi_2(v;d)=\sum_{n=0}^{\infty} c^{(2)}_n(d) v^n $ of the
 spin-1/2 Ising model with nearest-neighbor interaction,
 on a (hyper)-simple-cubical lattice
of general dimensionality $d$. For convenience, we have  reproduced here 
 also the expressions of the coefficients
$c^{(2)}_n(d)$ with $n<16$ that were already tabulated in
Ref.[\onlinecite{gofm}]  in terms of the variable $d$,
 whereas  here we have used 
the variable $q=2d$.  It should also be stressed
 that also in Ref.[\onlinecite{gofm}]
the series coefficients refer to the expansion variable $v$ like in this
Table and not to the variable $K$ as erroneously stated.}  \\ 
\endhead
\hline \multicolumn{1}{|r|}{{Continued on next page}}  \\ 
\hline
\endfoot
\hline 
\endlastfoot
 \hline
\scriptsize
$c_0^{(2)}=1$\\
$c_1^{(2)}=q$ \\
$c_2^{(2)}=q^2 - q$ \\
$c_3^{(2)}=q^3 - 2q^2 + q$ \\
$c_4^{(2)}=q^4 - 3q^3 + 2q^2 + q$ \\
$c_5^{(2)}=q^5 - 4q^4 + 4q^3 + 2q^2 - 3q$ \\
$c_6^{(2)}=q^6 - 5q^5 + 7q^4 - 2q^3 + 12q^2 - 23q$ \\
$c_7^{(2)}=q^7 - 6q^6 + 11q^5 - 8q^4 + 33q^3 - 88q^2 + 61q$ \\
$c_8^{(2)}=q^8 - 7q^7 + 16q^6 - 17q^5 + 33q^4 + 77q^3 - 652q^2 + 813q$ \\
$c_9^{(2)}=q^9 - 8q^8 + 22q^7 - 30q^6 + 41q^5 + 782/3q^4 - 5576/3q^3 
+ 10594/3q^2 - 6145/3q$ \\
$c_{10}^{(2)}=q^{10} - 9q^9 + 29q^8 - 48q^7 + 60q^6 + 595/3q^5 
+ 1132/3q^4 - 41404/3q^3 + 141125/3q^2 $\\$
- 45837q$ \\
$c_{11}^{(2)}=q^{11} - 10q^{10} + 37q^9 - 72q^8 + 94q^7 + 128q^6 
+ 2791q^5 - 112408/3q^4 + 140651q^3  $\\$  - 628610/3q^2 + 105297q$ \\
$c_{12}^{(2)}=q^{12} - 11q^{11} + 46q^{10} - 103q^9 + 148q^8 + 110/3q^7 
+ 7411/3q^6 - 4413q^5  $\\$ - 788360/3q^4 + 5431963/3q^3 - 13319981/3q^2 
+ 3721963q$ \\
$c_{13}^{(2)}=q^{13} - 12q^{12} + 56q^{11} - 142q^{10} + 228q^9 
- 278/3q^8 + 6760/3q^7 + 90088/3q^6  $\\$ - 2268283/3q^5 + 4828108q^4 
- 13659106q^3 + 52372552/3q^2 - 7921783q$ \\
$c_{14}^{(2)}=q^{14} - 13q^{13} + 67q^{12} - 190q^{11} + 341q^{10} 
- 282q^9 + 2163q^8 + 80084/3q^7 - 633892/3q^6  $\\$ - 14257841/3q^5 
+ 179124608/3q^4 - 269431329q^3 + 1643435774/3q^2 - 414746143q$ \\
$c_{15}^{(2)}=q^{15} - 14q^{14} + 79q^{13} - 248q^{12} + 495q^{11} 
- 1678/3q^{10} + 6724/3q^9 + 23674q^8  $\\$ + 347769q^7 - 78796706/5q^6 
+ 468273229/3q^5 - 2167816142/3q^4 + 5176873061/3q^3  $\\$ - 9973026754/5q^2 
+ 2545796651/3q$ \\
$c_{16}^{(2)}(d)=q^{16} - 15q^{15} + 92q^{14} - 317q^{13} + 699q^{12} 
- 2879/3q^{11} + 7663/3q^{10} $\\$  + 62404/3q^9 + 951902/3q^8 
- 29340047/5q^7 - 1222190629/15q^6 + 1842802906q^5 $\\$  - 40303287247/3q^4 
+ 727440333881/15q^3 - 436050363522/5q^2 + 61362596609q$ \\
$c_{17}^{(2)}(d)=q^{17} - 16q^{16} + 106q^{15} - 398q^{14} + 963q^{13} 
- 1526q^{12} + 3198q^{11} + 53594/3q^{10} $\\$  + 872213/3q^9 
+ 62822606/15q^8 - 1721035544/5q^7 + 24837980998/5q^6 $\\$ 
 - 103041998423/3q^5 
+ 658765700268/5q^4 - 1406657809016/5q^3 $\\$  + 1518632487582/5q^2 
- 124150140027q$ \\
$c_{18}^{(2)}(d)=q^{18} - 17q^{17} + 121q^{16} - 492q^{15} + 1298q^{14} 
- 6931/3q^{13} + 12916/3q^{12} $\\$  + 14580q^{11} + 803611/3q^{10} 
+ 58164043/15q^9 - 6630346477/45q^8 - 18614162023/15q^7  $\\$ 
+ 2490068122951/45q^6 - 9120432843283/15q^5+ 153498549866917/45q^4  $\\$ 
- 160125507535387/15q^3 
+ 791246681603369/45q^2 - 35100413743831/3q$ \\
$c_{19}^{(2)}(d)=q^{19} - 18q^{18} + 137q^{17} - 600q^{16} + 1716q^{15} 
- 10124/3q^{14} + 18163/3q^{13} $\\$  + 31678/3q^{12} + 746245/3q^{11} 
+ 10791224/3q^{10} + 155233339/3q^9 $\\$  - 119093401726/15q^8 
+ 797131628934/5q^7
 - 7738214717002/5q^6 $\\$  + 26108335469563/3q^5 
- 441756642770324/15q^4 + 870753648372433/15q^3
 $\\$ - 893908363283714/15q^2 + 23668071765201q$ \\
$c_{20}^{(2)}(d)=q^{20} - 19q^{19} + 154q^{18} - 723q^{17} 
+ 2230q^{16} - 4792q^{15} + 8677q^{14}  $\\$  + 15815/3q^{13} 
+ 701276/3q^{12} + 50129237/15q^{11} + 48537033q^{10} 
- 18335572678/5q^9 $\\$  - 109608216238/9q^8 + 8227897945731/5q^7 
- 1179114186108353/45q^6 $\\$  + 1059691462407483/5q^5 
- 9089741189180104/9q^4 + 2855597663272273q^3 $\\$  
- 65971006971414364/15q^2 + 2798133827599029q$ \\

\label{tab1}
\end{longtable}
\end{center}

\section{ High density expansion of the critical temperature}

By solving recursively the equation 
\begin{equation}
1/\chi_2(v_c;d)=0
\label{1suchi2}
\end{equation}
 with the Ansatz $v_c(d)= {\rm tanh}K_c(d)= a_1/q+a_2/q^2+...$ an
expansion of the critical temperature in inverse powers of $q$ was
computed in Ref.[\onlinecite{fishergaunt}] up to the fifth order. We
confirm the results of Ref.[\onlinecite{fishergaunt}], (which are
expressed in terms of the variable $K$ rather than the variable $v$
used above) and, taking advantage of our data, we are able to carry
the $1/q$ expansion to the 12th order:

\begin{eqnarray}
\nonumber \frac{1}{qK_c(d)}= 1 - \frac{1}{q} - \frac {4}{3q^{2}}
- \frac{13}{3q^{3}}-
       \frac{979 }{45q^{4}}
- \frac{2009}{15q^{5}}  -
       \frac{176749}{189q^{6}} 
- \frac{6648736 }{945q^{7}} 
  - \frac{765907148 }{14175q^{8}} \\ 
 - \frac{5446232381}{14175q^{9}}
- \frac{829271458256}{467775q^{10}} +
       \frac{164976684314}{22275q^{11}} 
  +\frac{6495334834824112} {638512875q^{12}}...
\label{kas}
\end{eqnarray}
A fifth order expansion of the same kind for $K_c(d)$ was also
 obtained\cite{gerberfisher} for the $n$-vector model.  All these
 expansions are presumably of asymptotic character, but so far this
 property has been established\cite{gerberfisher}
 only in the case of the spherical model,
 i.e. in the large $n$ limit.

\section{ Series Analyses}

We shall now very briefly
 discuss the numerical estimates of some non-universal
 critical parameters of the spin-1/2 Ising models for $d > 4$. In particular,
 we shall use the expansion of $\chi_2(K;d)$ to 
 locate  the critical points $K_c(d)$. We shall also estimate the critical
 amplitudes of the five lowest-order susceptibilities and a few
 universal ratios of these.  In our analysis, we
 have simply assumed that all (higher) susceptibilities show
 MF exponents of divergence, as also our recent
 work\cite{bp1,bp2} has contributed to confirm numerically.
 
  The critical parameters are defined by the asymptotic critical behaviors of
 the susceptibilities
\begin{equation}
\chi_{2p}(K;d) \approx A_{2p}(d) \tau(d)^{-\gamma_{2p}}\Big[1+a_{2p}(d)
\tau^{\theta(d)}...\Big]
\label{chi2pas}
\end{equation}
 as $K \rightarrow K_c(d)^-$.  Here $\tau(d)=(1 -K/K_c(d))$,
$\gamma_{2p}=\gamma_2+3(p-1)$ is the MF exponent, (assumed to
depend on the order $2p$ of the susceptibility, but not on the lattice
dimensionality for $d > 4$) and $\gamma_2=1$. 
$A_{2p}(d)$ denotes the amplitude of the leading
singularity, $a_{2p}(d)$ the amplitude of the leading correction to
scaling and $\theta(d)$ is the exponent of the leading
correction-to-scaling. The value of $\theta(d)$ is expected
\cite{guttm} to be $1/2$ for $d=5$, while for $d=6$ it
should be $1$ with a possible logarithmic multiplicative
correction. Generally, for $d>6 $ it is expected that
$\theta(d)=(d-4)/2$.

 We can only briefly outline the now standard numerical approximation
techniques that we have used for these analyses, since a more detailed
discussion was already given in Refs.[\onlinecite{bp1,bp2,bcisiesse}].
We have mainly employed the differential approximant(DA) method, that
generalizes\cite{guttda} the elementary well known Pad\'e approximant
method, to resum the HT expansions up to the border of their
convergence disks.  This technique estimates the values of the finite
quantities or the singularity parameters for the divergent quantities
from the solution, called {\it differential approximant}, of an
initial value problem for an ordinary linear inhomogeneous
differential equation of the first- or of a higher-order.  Various
differential equations can be formed from a given series
expansion. For each of them, the coefficients are polynomials in the
expansion variable, defined in such a way that the series expansion of
the solution of the equation equals, up to some appropriate order, the
series to be approximated.

Sometimes, to determine the location of the critical points, it is
more convenient to use a smoother and faster converging
modification\cite{zinnmra,guttda,bcisiesse}, called {\it
modified-ratio-approximant}(MRA) of the traditional methods of
extrapolation of the series coefficient-ratio-sequence.  The MRAs
produce sequences $(K_c^{(r)}(d))$ of approximations of the critical
point that can be easily extrapolated to large orders $r$ of expansion
and therefore in some cases they may yield more accurate estimates
than the DAs for which the analogous extrapolation is somewhat
arbitrary.  Let us finally stress that, when using the DAs the
evaluation of the uncertainties has not the same meaning as for MCs,
but remains subjective to some extent and only indicates  a small
multiple of the spread of the values of a conveniently large sample of
the highest-order approximants, formed from all or most expansion
coefficients. If the sample averages remain stable as the order of the 
 series grows and it can be assumed that stability indicates convergence,
 than the spread can be seen as a reasonable measure of the uncertainty
 of the results.

For the critical inverse temperatures $K_c(d)$ of the systems under
 study, we consider our best estimates those reported in Table
 \ref{tab2}. They are obtained from the HT expansion of the ordinary
 susceptibility $\chi_2(K;d)$ by extrapolating to large order of
 expansion, a few (from four to six) highest-order terms of the MRA
 sequence of estimates $(K^{(r)}_c(d))$ of the critical inverse
 temperature, basing  on a fit to their simple asymptotic
 behavior\cite{bcisiesse}
\begin{equation}
K^{(r)}_c(d)=K_c(d)- \frac{\Gamma(\gamma_2)} {\Gamma(\gamma_2-\theta)} 
\frac{\theta^2(1-\theta)a_2(d)}{r^{1+\theta}}+o(\frac{1}{r^{1+\theta}}).
\label{kras}
\end{equation}
A small multiple of the fit error is taken as the uncertainty of the
final estimate.  

In the case of six-dimensional lattices, we expect
$\theta=1$.  Therefore the second term on the right-hand side of
eq.(\ref{kras}) vanishes and it must be replaced by a higher-order
term reflecting the exponent of the next-to-leading correction to
scaling in eq.(\ref{chi2pas}).  In the $d=5$ case, in which one
expects $\theta=1/2$, the coefficient of $1/r^{1+\theta}$ in
eq.(\ref{kras}) appears to be numerically negligible, so that the
situation is similar to that of the six-dimensional case. In general,
to avoid making assumptions on the values of the exponents of the
next-to-leading correction to scaling, we have assumed an asymptotic
form $K^{(r)}_c=K_c+ w/r^{1+\epsilon}$ and determined $K_c$, $w$, and
the ``effective'' exponent $\epsilon=\epsilon(d)$ by a best fit to the
few highest-order terms of the sequence $(K^{(r)}_c)$.  We thus obtain
the values $\epsilon= 1.1(2)$ for $d=5$ and $\epsilon= 1.5(2)$ for
$d=6$. For $d>6$, the values of $\epsilon$ thus obtained are
larger. Therefore our estimates consistently confirm the above
mentioned expectations about $\theta(d)$, and suggest that in $d=5$
and $d=6$ the asymptotic behavior of eq.(\ref{kras}) is actually
determined by the next-to-leading, rather than the leading correction
to scaling.  On the other hand, for $d=5$ and $d=6$, a measure of the
exponent $\theta(d)$ of the leading corrections to scaling, whose
amplitudes $a_{2p}(d)$ are probably not negligible in spite of the
fact that they are not seen by the MRAs, can be performed studying by
DAs the critical behavior of appropriate universal ratios of higher
susceptibilities, such as those introduced below in
eqs.(\ref{Ii}),(\ref{Ai}) and (\ref{Bi}).  In these ratios the
dominant critical singularities cancel, while the leading corrections
to scaling generally survive and thus can be detected by DAs. This was
already discussed in Ref.[\onlinecite{bp2}]. In conclusion, all these
results are in reasonable agreement with the
expectations\cite{berche,guttm} indicated above.
\begin{table}[ht]\scriptsize
\caption{ Our estimates of the critical inverse temperatures $K_c(d)$,
 obtained from the ordinary susceptibility expansions,  
for several hsc lattices of dimensionality $d > 4$. 
 We have marked by an asterisk the estimates 
  in the cases in which expansions
 extend beyond the 20th order\cite{bp2}. In particular  
 for $d=5$ our series extend to the 22nd order, and for $d=6$ to 
the 21st order. } 
\begin{tabular}{|l|c|c|c|c|c|c|}
 \hline
 Source & $K_c(5)$ &$K_c(6)$ &$K_c(7)$ &$K_c(8)$ &$K_c(9)$ &$K_c(10)$\\
 \hline
HT This work &0.113920(1)*&0.092298(1)*&0.0777094(2)&
0.067155(1)&0.059148(1) &0.052858(1) \\
HT [\onlinecite{munkel,gofm}]  &0.113935(15)&0.092295(3)&0.077706(2)&&& \\
HT [\onlinecite{janke}]   & 0.113915(3)&&&&&\\
MC [\onlinecite{berche}]&0.113925(12)&0.092290(5)&0.077706(2)
&0.067144(4)&&\\
MC [\onlinecite{mer5,mer6}]&0.11391(?) &0.09229(4)&&&&\\
MC [\onlinecite{ak6,ak7,ak8}]&&0.09229(4)&0.0777(1)&0.06712(4)&&\\
MC [\onlinecite{luijten}]&0.1139152(4)&&&&&\\
MC [\onlinecite{lundow}] &0.1139139(5)&&&&& \\
 \hline
%\colrule 
 \end{tabular} 
 \label{tab2}
\end{table}

  In the Table \ref{tab2}, we have also reported a few of the most
  recent estimates of the critical inverse temperatures obtained in
  the literature either from shorter HT series or from MC simulations,
  for the values of $d$ considered in our study.  Since, for $d>4$, no
  logarithmic factors are expected to modify the leading MF behavior
  of the physical quantities, our series analyses are likely to yield
  estimates of a high accuracy, that moreover seem to improve with
  increasing lattice dimensionality, both because of the decreasing
  influence of the corrections to scaling and of the increasing
  lattice coordination number. All the results obtained from MRAs are
  consistent, within their uncertainties, also with the analyses
  employing DAs.  In $d=5$ dimensions, our estimate is slightly larger
  than other estimates\cite{luijten,lundow} of similar nominal
  accuracy, but can be  considered essentially compatible with those of
  Refs.[\onlinecite{berche,gofm,munkel,janke}].  It is of interest to
  quote here also our estimate $K_c(5)=0.113919(2)$ obtained from second-order
  quasi-diagonal DAs that use all series coefficients up to order $20
  \leq l \leq 22$. The same value, with a slightly larger uncertainty,
  is obtained from DAs using all series coefficients up to order $18 \leq
  l \leq 20$.  Generally, for higher values of $d$, 
  our estimates of the critical inverse temperatures do
  not differ much from the old values, but they show a greater
  accuracy. For $d>7$, the estimates of $K_c(d)$ obtained from the $1/d$
  expansion of eq.(\ref{kas}) reproduce our MRA values 
  within the errors.

 Let us now turn to the critical amplitudes $A_{2p}(d)$ 
of the susceptibilities $\chi_{2p}(K;d)$ with $n=1,2,...5$, that 
can be determined, in terms of the previously estimated values of $K_c(d)$, 
by extrapolating the effective amplitudes

\begin{equation}
A^{eff}_{2p}(K;d)= (1-K/K_c(d))^{\gamma_{2p}} \chi_{2p}(K;d)
\label{ampef}
\end{equation}

 to $K=K_c(d)$, namely $A_{2p}(d)=A^{eff}_{2p}(K_c;d)$.
 Our analysis uses first- and 
second-order DAs of the HT expansion of $A^{eff}_{2p}(K;d)$. 

\begin{table}[ht]\scriptsize
\caption{ Our estimates of the critical amplitudes $A_{2p}(d)$ of the
  susceptibilities $\chi_{2p}(K;d)$, normalized to their values 
$A^{MF}_{2p}$ in the 
 MF approximation for
 several hyper-simple-cubical lattices of dimensionality $d > 4$.  We
 have marked by an asterisk the estimates obtained from series
 extending beyond the 20th order. (For $d=5$ our series extend to the 22nd
 order, and for $d=6$ the 21st order.) }
\begin{tabular}{|c|c|c|c|c|c|c|c|}
 \hline
Amplitude& Source & $d=5$ &$d=6$ &$d=7$ &$d=8$ &$d=9$ &$d=10$    \\
 \hline
$A_2(d)/A_2^{MF}$& This work &1.32(1)*&1.179(2)*&1.124(2)&
1.096(2)&1.079(1) &1.067(3) \\
$A_4(d)/A_4^{MF}$& This work &1.40(1)*&1.20(1)*&1.138(2)&
1.105(2)&1.085(1) &1.068(3) \\
$A_6(d)/A_6^{MF}$& This work &1.49(1)*&1.22(1)*&1.147(3)&
1.114(2)&1.091(2) &1.077(3)\\
$A_8(d)/A_8^{MF}$&This work&1.57(1)*&1.25(1)*&1.165(2)&1.122(2)
&1.097(3)&1.081(4)\\ 
$A_{10}(d)/A_{10}^{MF}$&This work&1.65(2)*&1.28(1)*&1.19(1)&1.13(1)
&1.103(4)&1.085(4)\\ 
$A_2(d)/A_2^{MF}$ &MC [\onlinecite{berche}]&1.291(3)&1.1606(17)&1.1008(5)
&1.0836(5)&&\\
$A_2(d)A_2^{MF}$& HT [\onlinecite{guttm}]&1.311(9)&1.168(8)&&&&\\
 \hline
%\colrule 
 \end{tabular} 
 \label{tab3}
\end{table}
In the Table \ref{tab3}, we have reported our series estimates of
$A_{2}(d)$, $A_{4}(d)$, $A_{6}(d)$, $A_8(d)$ and $A_{10}(d)$
normalized to their values in the MF approximation\cite{joyce}:
 $A_2^{MF}=1$, $A_4^{MF}=-2$, $A_6^{MF}=40$,
$A_8^{MF}=-2240$ and $A_{10}^{MF}=246400$. 
As expected, these ratios tend to unity as
$d \rightarrow \infty$. For comparison, we have also reported the
corresponding MC results of Ref.[\onlinecite{berche}].  The results of this 
simulation are slightly, but systematically smaller than our HT
estimates.  This minor disagreement cannot be due to the very small
difference in the estimates of $K_c(d)$ used in the two calculations,
but must probably be ascribed to an underestimate of the uncertainties
inherent in the MC generation of the HT series.  On the other hand,
the old estimates\cite{guttm} of the same amplitudes obtained from an
analysis of the 11th order HT series of Ref.[\onlinecite{fishergaunt}]
are compatible with ours.  Estimates for $A_{2}(d)$ have also been
obtained\cite{gaunt}  from a third order expansion in $1/d$, but
they are not accurate enough.  No comparison at all was possible for
$A_{4}(d)$,...,$A_{10}(d)$ for which we do not know of other estimates
in the literature.

 Finally, we have computed also for $d>6$, the critical values of a
 few universal ratios of higher susceptibilities such as the lowest
 order terms in the sequences ${\cal I}^+_{2r+4}(d)$, ${\cal A}^+_{2r+4}(d)$
 and ${\cal B}^+_{2r+8}(d)$, defined by

\begin{equation}
{\cal I}^+_{2r+4}(d)
=\lim_{K \to K^-_c}\frac{\chi_2(K;d)^r \chi_{2r+4}(K;d)} 
{\chi_{4}(K;d)^{2r+1}}=\frac{A_2(d)^r A_{2r+4}(d)} {A_4(d)^{2r+1}}
\label{Ii} 
\end{equation}
\begin{equation} 
{\cal A}^+_{2r+4}(d)=\lim_{K \to K^-_c}\frac{\chi_{2r}(K;d) 
\chi_{2r+4}(K;d)}{\chi_{2r+2}(K;d)^2}
=\frac{A_{2r}(d)A_{2r+4}(d)} {A_{2r+2}(d)^2} 
\label{Ai} 
\end{equation}
 \begin{equation}
 {\cal B}^+_{2r+8}(d)=\lim_{K \to K^-_c}\frac{\chi_{2r}(K;d) 
\chi_{2r+8}(K;d)}{\chi_{2r+4}(K;d)^2}
=\frac{A_{2r}(d)A_{2r+8}(d)} {A_{2r+4}(d)^2}
\label{Bi} 
\end{equation}
 for $r > 0$.
These universal ratios were introduced in Ref.[\onlinecite{watson}] and were
 studied in detail  for $d=4,5,6$  in Ref.[ \onlinecite{bp1,bp2}].  
For the first few values of $r=1,2,3$, 
we have checked that as expected, also for $d>6$, they   
take  the MF values: 
  ${\cal I}_{6}^{+MF}=10$, 
${\cal I}_{8}^{+MF}=280$, 
${\cal I}_{10}^{+MF}=15400$, ${\cal A}_{8}^{+MF}=14/5$,  
${\cal A}_{10}^{+MF}=55/28$  and ${\cal B}_{10}^{+MF}=154$
  within a relative accuracy generally 
higher than $10^{-3}$, although
  the single amplitudes entering into the ratios reach their MF
 value only in the $d \rightarrow \infty$ limit.

\section{Conclusions}

We have represented in a compact and exact form, as simple polynomials in the
lattice dimensionality, the HT expansion coefficients of the (higher)
susceptibilities in the case of the spin-1/2 Ising model on the hsc
lattices of general dimensionality $d$.  Our calculations add five
more orders to the existing expansions of the ordinary susceptibility
for general $d$ and nine more orders to those of the fourth- and
sixth-order susceptibilities. For the susceptibilities of order
greater than the sixth no such data already exist in the literature.

An analysis of the series for lattice dimensionality $d>4$ yields
estimates of non-universal parameters that compare well with the
previous results whenever available, but are generally more accurate.
Our estimates of a few universal ratio amplitudes, provide a high
accuracy check that, unsurprisingly they take MF values for $d>4$.

Finally, using the general $d$ expression of the ordinary
susceptibility, we have been able to expand up to the 12th order the
critical temperature in powers of $1/d$ more than doubling the length
of the  expansion already known in the literature.

\section{Acknowledgements}
  We thank the Physics Departments of Milano-Bicocca University and of
Milano University for hospitality and support.  Partial support by the
MIUR is also acknowledged.

\section{Appendix}

\subsection{The HT  expansion of $\chi_4(v;d)$}

\begin{center}
\begin{longtable}{l}
\caption {The coefficients $c^{(4)}_n(d)$ of the HT series expansion in
powers of $v={\rm tanh}K$ for the  susceptibility
$\chi_4(v;d)=\sum_{n=0}^{\infty} c^{(4)}_n(d) v^n $ of the 
spin-1/2 Ising model on a (hyper)-simple-cubical lattice of 
dimensionality $d$ as polynomials in the coordination number $q=2d$.}
 \\
\endhead
\hline \multicolumn{1}{|r|}{{Continued on next page}} \\ 
\hline
\endfoot
\hline
\endlastfoot
\hline
$c_0^{(4)}=-2$\\
$c_1^{(4)}=-8q$ \\
$c_2^{(4)}=-20q^2 + 14q$ \\
$c_3^{(4)}=-40q^3 + 64q^2 - 24q$ \\
$c_4^{(4)}=-70q^4 + 180q^3 - 122q^2 - 14q$ \\
$c_5^{(4)}=-112q^5 + 400q^4 - 408q^3 - 48q^2 + 168q$ \\
$c_6^{(4)}=-168q^6 + 770q^5 - 1080q^4 + 152q^3 + 146q^2 + 578q$ \\
$c_7^{(4)}=-240q^7 + 1344q^6 - 2440q^5 + 1120q^4 - 1040q^3 + 5360q^2 
- 4424q$ \\
$c_8^{(4)}=-330q^8 + 2184q^7 - 4914q^6 + 3850q^5 - 3888q^4 + 8464q^3 
+ 14374q^2 - 33774q$ \\
$c_9^{(4)}=-440q^9 + 3360q^8 - 9072q^7 + 10016q^6 - 10144q^5 
+ 2864/3q^4 + 408256/3q^3 - 980528/3q^2 $\\$ + 618536/3q$ \\
$c_{10}^{(4)}=-572q^{10} + 4950q^9 - 15648q^8 + 22232q^7 - 23722q^6 -
 34580/3q^5 + 763630/3q^4 - 37060/3q^3  $\\$ - 6036988/3q^2 + 2586414q$ \\
$c_{11}^{(4)}=-728q^{11} + 7040q^{10} - 25560q^9 + 44352q^8 - 52024q^7 
- 19616q^6 + 221136q^5 + 7971968/3q^4  $\\$ - 15033816q^3 + 76181488/3q^2 
- 13485336q$ \\
$c_{12}^{(4)}=-910q^{12} + 9724q^{11} - 39930q^{10} + 81810q^9 
- 107730q^8 - 18080/3q^7 + 182096/3q^6 $\\$  + 17071112/3q^5 
- 42028180/3q^4 - 62677128q^3 + 806882456/3q^2 - 262544882q$ \\
$c_{13}^{(4)}=-1120q^{13} + 13104q^{12} - 60104q^{11} + 142000q^{10} 
- 211128q^9 + 182912/3q^8 - 668440/3q^7 $\\$+ 19780304/3q^6   
+ 41895984q^5 - 1655518208/3q^4 + 5642509000/3q^3 - 7772420336/3q^2 $ \\$
+ 1218754376q$ \\
$c_{14}^{(4)}=-1360q^{14} + 17290q^{13} - 87672q^{12} + 234696q^{11} 
- 393054q^{10} + 237158q^9 - 661318q^8 $\\$+ 17401420/3q^7 + 338427338/3q^6 
- 740356150q^5 - 2254648078/3q^4 + 47003116718/3q^3 $\\$- 125022788422/3q^2 
+ 34659398258q$ \\
$c_{15}^{(4)}=-1632q^{15} + 22400q^{14} - 124488q^{13} + 372512q^{12} 
- 698512q^{11} + 615952q^{10} - 1348840q^9 $\\$+ 3731152q^8 + 454575800/3q^7 
+ 5824320064/15q^6 - 17623120168q^5 + 322919474240/3q^4 $\\$
- 857689960864/3q^3 + 5197801864336/15q^2 - 452601849976/3q$ \\
$c_{16}^{(4)}=-1938q^{16} + 28560q^{15} - 172690q^{14} + 571402q^{13} 
- 1191044q^{12} + 1346952q^{11} - 2483954q^{10} $\\$+ 2574458/3q^9 
+ 163363530q^8 + 5936891456/3q^7 - 427652053304/15q^6 + 49753283730q^5$\\$ +
 1888295056504/3q^4 - 11165804886700/3q^3 + 39268765896758/5q^2 
- 5885730502990q$ \\
$c_{17}^{(4)}=-2280q^{17} + 35904q^{16} - 234720q^{15} + 851200q^{14} 
- 1957920q^{13} + 2660560q^{12} - 4434752q^{11} $\\$- 2180496q^{10} 
+ 458206840/3q^9 + 9171471680/3q^8 - 96727802776/15q^7 
- 7627486265248/15q^6 $\\$+ 15874774585472/3q^5 - 23613288071104q^4 
+ 809239034177296/15q^3 - 300645818028224/5q^2 $\\$+ 24960425046264q$ \\
$c_{18}^{(4)}=-2660q^{18} + 44574q^{17} - 313344q^{16} + 1236200q^{15} 
- 3116218q^{14} + 14699672/3q^{13} $\\$- - 23502482/3q^{12} - 13194884/3q^{11} 
+ 369135052/3q^{10} + 11095980830/3q^9 + 260264951278/9q^8 
 $\\$-- 14453314408166/15q^7 + 218881556715946/45q^6 + 45588473570314/3q^5 
- 2179954835300908/9q^4  $\\$ + 14858869355291746/15q^3 
- 81689671393093936/45q^2 + 3792619258364714/3q$ \\
$c_{19}^{(4)}=-3080q^{19} + 54720q^{18} - 411672q^{17} + 1755776q^{16} 
- 4819864q^{15} + 25683904/3q^{14} - 41145344/3q^{13} $\\$- 12513184/3q^{12} 
+ 229186600/3q^{11} + 11910644656/3q^{10} + 55759673624q^9 
- 1666280612320/3q^8 $\\$- 199407718261024/15q^7 + 1180721878106528/5q^6 
- 4913912212688992/3q^5$\\$ + 18295751469102064/3q^4 - 188910340455291256/15q^3
 + 198430901982546976/15q^2$\\$ - 5313569529072792q$ \\
$c_{20}^{(4)}=-3542q^{20} + 66500q^{19} - 533178q^{18} + 2445042q^{17} 
- 7267702q^{16} + 14344984q^{15}$\\$ - 23696940q^{14} + 3616100/3q^{13} 
+ 12996288q^{12} + 58969981544/15q^{11} + 75196713528q^{10} 
$\\$+ 232751024986q^9 - 1367841862391708/45q^8 + 1437869303984316/5q^7 
$\\$- 10204868446463528/45q^6 - 12244568432836332q^5 
+ 4116746384585661118/45q^4 $\\$- 1535126389650383104/5q^3 
+ 7658022498562556906/15q^2 - 336280716221910846q$ \\

\label{tab4}
\end{longtable}
\end{center}

\subsection{The HT  expansion of $\chi_6(v;d)$}

\begin{center}
\begin{longtable}{l}
\caption{ The coefficients $c^{(6)}_n(d)$ of the HT series expansion in
 powers of $v={\rm tanh}K$ for the  susceptibility
 $\chi_6(v;d)=\sum_{n=0,\infty} c^{(6)}_n(d) v^n $ of the 
 spin-1/2 Ising model on a (hyper)-simple-cubical lattice of 
 dimensionality $d$ as polynomials in $q = 2d$.}\\ 
\endhead
\hline \multicolumn{1}{|r|}{{Continued on next page}} \\ 
\hline
\endfoot
\hline
\endlastfoot
$c_0^{(6)}=16$\\ 
$c_1^{(6)}=136q$ \\
 $c_2^{(6)}=616q^2 - 376q$\\ 
$c_3^{(6)}=2016q^3 - 2912q^2 + 1016q$ \\
 $c_4^{(6)}=5376q^4 - 12768q^3 + 8592q^2 + 16q$ \\ 
$c_5^{(6)}=12432q^5
 - 41664q^4 + 42584q^3 - 1568q^2 - 11784q$ \\ 
$c_6^{(6)}=25872q^6 -
 112560q^5 + 157752q^4 - 33552q^3 - 48064q^2 - 16488q$ \\
 $c_7^{(6)}=49632q^7 - 266112q^6 +480816q^5 - 221888q^4 - 37848q^3 -
 380032q^2 + 409096q$ \\ 
$c_8^{(6)}=89232q^8 - 569184q^7 + 1270752q^6
 - 944832q^5 + 292432q^4 - 1414576q^3 + 364752q^2$ \\ $ + 2177328q$ \\
 $c_9^{(6)}=152152q^9 - 1125696q^8 + 3010224q^7 - 3132864q^6 +
 1795080q^5 - 7977904/3q^4$ \\ $ -32649872/3q^3 + 112890736/3q^2 -
 78703720/3q$ \\ 
$c_{10}^{(6)}=248248q^{10} - 2090088q^9 + 6537168q^8
 - 8795136q^7 + 6811560q^6 - 10886696/3q^5$ \\ 
$ - 139989920/3q^4 + 294356480/3q^3 +283502264/3q^2 - 233678136q$ \\
 $c_{11}^{(6)}=390208q^{11} - 3683680q^{10} + 13226664q^9 -
 21879744q^8 + 21066864q^7 - 5828736q^6$\\ 
$ -106751176q^5 - 245869072/3q^4 + 1897397904q^3 - 11488470128/3q^2
+ 2172800952q$ \\
 $c_{12}^{(6)}=594048q^{12} - 6214208q^{11} + 25234352q^{10} -
 49604016q^9 + 57229344q^8 - 17574592q^7$ \\ 
$ -177848112q^6 - 2536571600/3q^5 + 5835827312q^4 - 8921212960/3q^3 
- 23548883584q^2 $\\$  + 30251052208q$ \\ 
$c_{13}^{(6)}=879648q^{13} - 10098816q^{12} +
 45813768q^{11} - 104423264q^{10} + 141116904q^9 - 64357856q^8$ \\ $ -
 228295552q^7 - 2298496240q^6 + 16344675328/3q^5 + 200049349136/3q^4 -
 946502181448/3q^3$ \\ $ + 1442426211616/3q^2 -236344803688q$ \\
 $c_{14}^{(6)}=1271328q^{14} - 15890784q^{13} + 79721096q^{12} -
 206842064q^{11} + 321905760q^{10} $\\$  - 209254488q^9 - 189367680q^8
 - 4356596032q^7 - 24934282288/3q^6 + 731828251760/3q^5 $\\$  -
 1884079942960/3q^4 - 2540774076440/3q^3 + 14787583499552/3q^2 -
 4788568201288q$ \\ 
$c_{15}^{(6)}=1798464q^{15} - 24310272q^{14} +
 133721952q^{13} - 389311104q^{12} + 688157624q^{11} - 590354816q^{10} $\\$ 
 + 91288088q^9 - 6810138432q^8 - 120995015840/3q^7 +
 5693635677872/15q^6 + 4530618613376/3q^5 $\\$  - 56128080444176/3q^4 +
 175625859309352/3q^3 - 1134460813568032/15q^2 + 101715603123320/3q$
 \\ $c_{16}^{(6)}=2496144q^{16} - 36279360q^{15} + 217215936q^{14} -
 701494976q^{13} + 1392033552q^{12} $\\$   - 1485565136q^{11} +
 949304208q^{10} - 9440366096q^9 - 276020877632/3q^8 +
 3692327961344/15q^7 $\\$   + 40259238539312/5q^6 - 145655300075488/3q^5 +
 117263299297904/3q^4 + 1915937192446272/5q^3 $\\$   - 5668662187831232/5q^2
 + 939412612077584q$ \\ 
$c_{17}^{(6)}=3405864q^{17} - 52961664q^{16} +
 342995808q^{15} - 1217241984q^{14} + 2684716216q^{13} $\\$  -
 3419593264q^{12} + 3097994592q^{11} - 12255126320q^{10} -
 161764858632q^9 - 1523657245328/5q^8 $\\$  + 234835462900528/15q^7 +
 7119519453424/15q^6 - 867980933559704q^5 + 76620462792781184/15q^4 $\\$  -
 193297122994892848/15q^3 + 74970873477425792/5q^2 -
 6350764077866232q$ \\ 
$c_{18}^{(6)}=4576264q^{18} - 75806808q^{17} +
 528159264q^{16} - 2043636096q^{15} + 4966825416q^{14} $\\$  -
 22001068024/3q^{13} + 23949277744/3q^{12} - 47908848800/3q^{11} -
 735961413304/3q^{10} $\\$  - 6769987372664/5q^9 + 836907198003728/45q^8 +
 1061178583645504/5q^7  $\\$ - 123680489462001464/45q^6 +
 128471706713417032/15q^5 + 557653654570467832/45q^4 $\\$  -
 2046057519029273912/15q^3 + 13722905419902781064/45q^2 -
 683992228275224776/3q$ \\ 
$c_{19}^{(6)}=6063904q^{19} -
 106600032q^{18} + 795192408q^{17} - 3332563584q^{16} +
 8858492640q^{15} $\\$  - 44537998400/3q^{14} + 55191538840/3q^{13} -
 68853616240/3q^{12} - 1006105970464/3q^{11} $\\$  - 2925741243056q^{10} +
 38101102223000/3q^9 + 2578706022827728/5q^8 - 34953325549594928/15q^7 $\\$ 
 - 469486142941733296/15q^6 + 1067819443550786792/3q^5 -
 7740457369453181728/5q^4 $\\$  + 17085072081639612176/5q^3 -
 55533128504820921824/15q^2 + 1509351684072604536q$ \\
 $c_{20}^{(6)}=7934080q^{20} - 147517216q^{19} + 1173245136q^{18} -
 5295282672q^{17} + 15294767808q^{16} $\\$  - 28639450752q^{15} +
 39499504784q^{14} - 116974101616/3q^{13} - 420148419632q^{12} $\\$  -
 74964569272096/15q^{11} - 4879661509088q^{10} + 3832420365724432/5q^9 $\\$ 
 + 33799538542615616/9q^8 - 649819887125975392/5q^7 +
 36485867573612822272/45q^6 $\\$  - 5101763740998382112/5q^5 -
 89646139040158737760/9q^4 + 255768024175232201184/5q^3 $\\$  -
 1458409290472901242384/15q^2 + 67555026246347718672q$ \\
\label{tab6}
\end{longtable}
\end{center}

\subsection{The  HT expansion of $\chi_8(v;d)$}

\begin{center}
\begin{longtable}{l}
\caption{ The coefficients $c^{(8)}_n(d)$ of the HT series expansion in
 powers of $v={\rm tanh}K$ for the  susceptibility
 $\chi_8(v;d)=\sum_{n=0,\infty} c^{(8)}_n(d) v^n $ of the 
 spin-1/2 Ising model on a (hyper)-simple-cubical lattice of 
 dimensionality $d$ as polynomials in $q = 2d$.}\\
\endhead
\hline \multicolumn{1}{|r|}{{Continued on next page}} \\ 
\hline
\endfoot
\hline
\endlastfoot
\hline
$c_0^{(8)}=-272$\\ 
$c_1^{(8)}=-3968q$ \\ 
$c_2^{(8)}=-28160q^2
+ 16064q$ \\ 
$c_3^{(8)}=-135680q^3 + 185344q^2 - 63104q$ \\
$c_4^{(8)}=-508640q^4 + 1152960q^3 - 774352q^2 + 39376q$ \\
$c_5^{(8)}=-1595264q^5 + 5139200q^4 - 5262848q^3 + 626432q^2 +
1092480q$ \\ 
$c_6^{(8)}=-4374656q^6 + 18399040q^5 - 25800320q^4 +
7123072q^3 + 7771104q^2 - 451712q$ \\ 
$c_7^{(8)}=-10799360q^7 +
56229888q^6 - 101418880q^5 + 50302720q^4 + 20789760q^3 + 28718336q^2 $\\$  -
48450944q$ \\ 
$c_8^{(8)}=-24493040q^8 + 152280128q^7 - 338701088q^6 +
255076000q^5 - 7388208q^4 + 191935520q^3 $\\$  - 212236432q^2 - 180336624q$
\\ $c_9^{(8)}=-51800320q^9 + 374682880q^8 - 996706304q^7 +
1028397312q^6 - 349984384q^5  $\\$ + 1895565824/3q^4 + 2459602432/3q^3 -
16331584256/3q^2 + 12754084736/3q$ \\ 
$c_{10}^{(8)}=-103307776q^{10} +
852508800q^9 - 2649851776q^8 + 3506874624q^7 - 2074473600q^6 $\\$  +
4719123136/3q^5 + 24219243904/3q^4 - 83386258816/3q^3 +
12799466528/3q^2 + 28516712832q$ \\ 
$c_{11}^{(8)}=-195979264q^{11} +
1817221120q^{10} - 6480531200q^9 + 10519474944q^8 - 8601005312q^7  $\\$ +
4050302720q^6 + 31146711936q^5 - 146794482688/3q^4 - 276543258112q^3 +
2167133336320/3q^2 $\\$  - 442103339136q$ \\ 
$c_{12}^{(8)}=-356075200q^{12} +
3665247872q^{11} - 14777517040q^{10} + 28499934320q^9 - 29492323520q^8
 $\\$ + 39123812096/3q^7 + 248923396192/3q^6 + 38629503424q^5 -
4714014444080/3q^4 $\\$  + 7509887769376/3q^3 + 6420043301440/3q^2 -
4809699889232q$ \\ 
$c_{13}^{(8)}=-623054080q^{13} + 7049482752q^{12} -
31748068352q^{11} + 71050636800q^{10} - 88843430016q^9  $\\$ +
141501441280/3q^8 + 520697373184/3q^7 + 1366919398144/3q^6 -
12055727532544/3q^5 $\\$  - 6493984716800q^4 + 64312423603200q^3 -
335264513674496/3q^2 + 57772980342656q$ \\
$c_{14}^{(8)}=-1054685440q^{14} + 13009509760q^{13} - 64792589056q^{12}
+ 165243515008q^{11} - 242500649248q^{10} $\\$  + 163985982336q^9 +
289429527488q^8 + 4584901860992/3q^7 - 16199812722112/3q^6 $\\$  -
184263912218624/3q^5 + 822200412947456/3q^4 - 153864094174240q^3 -
2216382891425920/3q^2 $\\$  + 922208891532608q$ \\
$c_{15}^{(8)}=-1733636608q^{15} + 23152640000q^{14} - 126442211584q^{13}
+ 362278232576q^{12} - 611273245440q^{11} $\\$  + 1556604583424/3q^{10} +
1003359499648/3q^9 + 3663324892416q^8 - 1635463424q^7  $\\$ -
969658650984192/5q^6 + 705546067707904/3q^5 + 10957704133482496/3q^4 -
44381419834132480/3q^3 $\\$  + 103569486797574912/5q^2 -
28828949928421504/3q$ \\ $
c_{16}^{(8)}=-2775830992q^{16} +
39900474240q^{15} - 237223499840q^{14} + 754843908416q^{13} -
1441747180848q^{12} $\\$  + 4473073125664/3q^{11} - 69911950064/3q^{10} +
22061233174384/3q^9 + 65837014833920/3q^8 $\\$  - 384138748886400q^7 -
21910654023360544/15q^6 + 18929575971622144q^5 -
134319711356791312/3q^4 $\\$  - 78918132289769312/3q^3 +
1092518218137459408/5q^2 - 210107883695368240q$ \\
$c_{17}^{(8)}=-4340913280q^{17} + 66818684416q^{16} - 429792378880q^{15}
+ 1504494886400q^{14} - 3212833578240q^{13} $\\$  + 3936047152128q^{12} -
1715109963776q^{11} + 40032013286144/3q^{10} + 222805972125440/3q^9  $\\$ -
1579582584876032/3q^8 - 32879942504028672/5q^7 +
181099743084921344/5q^6 $\\$  + 356398939022048512/3q^5 -
1318003726217876480q^4 + 19130142996335340032/5q^3 $\\$  -
23490140191433877504/5q^2 + 2038219551714236544q$ \\
$c_{18}^{(8)}=-6645196800q^{18} + 109051063616q^{17} -
754773335296q^{16} + 2883607564800q^{15} $\\$  - 6815490996224q^{14} +
28995994718080/3q^{13} - 21216015570880/3q^{12} +
69781088168704/3q^{11} $\\$  + 174184899405984q^{10}- 1227284831435200/3q^9
- 143067775267929728/9q^8 + 84830862839461504/15q^7 $\\$  +
43761296741698393216/45q^6 - 5137578702684892480q^5 +
57115043408829771200/9q^4 $\\$  + 314661275511421183136/15q^3 -
3143073756109544998336/45q^2 + 173286428915305918784/3q$ \\
$c_{19}^{(8)}=-9977510400q^{19} + 173882649600q^{18} -
1288851943808q^{17} + 5338095665664q^{16} $\\$  - 13846023547392q^{15} +
66947182003712/3q^{14} - 64440145279744/3q^{13} + 41372635858944q^{12} $\\$ 
+ 340483735795968q^{11} + 885961624852736/3q^{10} -
84604254215282432/3q^9 - 148038049578011648q^8  $\\$ +
37110053873309522176/15q^7 - 22502517113219316736/15q^6 -
253892478849624551680/3q^5 $\\$  + 1436655318187681722368/3q^4 -
17346420241533257325056/15q^3 + 19533815567719217760256/15q^2 $\\$  -
540509926226296804992q$ \\ 
$c_{20}^{(8)}=-14718406560q^{20} +
271460904000q^{19} - 2145801037072q^{18} + 9579095190288q^{17} $\\$  -
27071038685440q^{16} + 48866734786048q^{15} - 56625604724256q^{14} +
236514693001600/3q^{13} $\\$  + 1767533919116336/3q^{12} +
30408195156033568/15q^{11} - 39513935844411504q^{10} $\\$  -
488739512196072000q^9 + 137400352859917843312/45q^8 +
112018250037801599872/3q^7 $\\$  - 18735633799569055366448/45q^6 +
4151594372466586494272/3q^5 - 4171180399934389754912/45q^4  $\\$ -
53306960347891458594336/5q^3 + 382608439113501754024976/15q^2 -
19084733342231658023472q$ \\
\label{tab8} 
\end{longtable}
\end{center}

\subsection{The  HT expansion of $\chi_{10}(v;d)$}

\begin{center}
\begin{longtable}{l}
\caption{ The coefficients $c^{(10)}_n(d)$ of the HT series expansion in
 powers of $v={\rm tanh}K$ for the  susceptibility
 $\chi_{10}(v;d)=\sum_{n=0,\infty} c^{(10)}_n(d) v^n $ of the 
 spin-1/2 Ising model on a (hyper)-simple-cubical lattice of 
 dimensionality $d$ as polynomials in $q = 2d$.} \\
\endhead
\hline \multicolumn{1}{|r|}{{Continued on next page}} \\ 
\hline
\endfoot
\hline
\endlastfoot
\hline

 $c_0^{(10)}=7936$\\ 
$c_1^{(10)}=176896q$ \\
$c_2^{(10)}=1805056q^2 - 990976q$ \\
 $c_3^{(10)}=11977856q^3 -
15817472q^2 + 5331456q$ \\
 $c_4^{(10)}=59835776q^4 - 131649408q^3 +
88440832q^2 - 7113344q$ \\ 
$c_5^{(10)}=243571328q^5 - 764301824q^4 +
784337664q^3 - 132473088q^2 - 130771200q$ \\ $c_6^{(10)}=848335488q^6
- 3485722240q^5 + 4893727872q^4 - 1578948352q^3 - 1335948160q^2 +
306159872q$ \\ $c_7^{(10)}=2611834368q^7 - 13319674368q^6 +
24015744896q^5 - 12725984768q^4 - 5562137984q^3$\\ $ - 1363634176q^2 +
7148102656q$ \\ $c_8^{(10)}=7273863168q^8 - 44390201856q^7 +
98556826368q^6 - 76197347072q^5 - 5531303424q^4$\\ $ - 23187248384q^3 +
55450859776q^2 + 16088832768q$ \\ $c_9^{(10)}=18637718528q^9 -
132572504064q^8 + 351628333056q^7 - 365569735168q^6 + 83564687744q^5$\\$ -
362811788800/3q^4 - 13891630208/3q^3 + 2898002725888/3q^2 -
2566639813120/3q$ \\ 
$c_{10}^{(10)}=44506981376q^{10} - 361758938112q^9
+ 1120211169792q^8 - 1480274059264q^7 + 724243422848q^6$\\$ -
1363066769792/3q^5 - 4514271066752/3q^4 + 22869242557184/3q^3$\\$ -
13265911387264/3q^2 - 4264749231360q$ \\
$c_{11}^{(10)}=100054712576q^{11} - 915089077760q^{10} +
3249120416256q^9 - 5242545724416q^8 + 3817905779072q^7$\\$ -
1699281827840q^6 - 9196873191808q^5 + 81171677099264/3q^4 +
41908080196608q^3$\\$ - 498024823056896/3q^2 + 110610239423232q$ \\
$c_{12}^{(10)}=213450555136q^{12} - 2169762639616q^{11} +
8706410938368q^{10} - 16654558212864q^9$\\$ + 15937418585088q^8 -
21327477729536/3q^7 - 104234607142528/3q^6 + 147562279746944/3q^5$\\$ +
1278777987933440/3q^4 - 1015433727966592q^3 + 18678581451392/3q^2 +
969932534412928q$ \\ 
$c_{13}^{(10)}=434934497536q^{13} -
4864839662592q^{12} + 21799836371456q^{11} - 48343942806016q^{10}$\\$ +
57122586837888q^9 - 91561796736256/3q^8 - 298714168559488/3q^7 +
6166179994112/3q^6$\\$ + 1764241975937664q^5 - 2494450072249088/3q^4 -
46019039628909440/3q^3 $\\$ + 94924943316071680/3q^2 - 17299176815162368q$
\\ $c_{14}^{(10)}=851007283456q^{14} - 10386976407808q^{13} +
51466387045632q^{12} - 130025893949952q^{11}$\\$ + 182747477058432q^{10} -
122315166083968q^9 - 228199423744512q^8 - 954749235840512/3q^7$\\$ +
13606511889524864/3q^6 + 13727427846676864q^5 -
317929887704789632/3q^4 $\\$+ 396942583026124160/3q^3 +
348318619589400320/3q^2 - 227789642969636608q$ \\
$c_{15}^{(10)}=1606033313280q^{15} - 21240894101504q^{14} +
115402828579584q^{13} - 327562821168128q^{12}$\\$ + 533995375623424q^{11}
- 443300013544960q^{10} - 401964295244160q^9 - 1341116402467072q^8$\\$ +
23002048231205120/3q^7 + 1170205670425126912/15q^6 -
284186929185738880q^5$\\$  -  2073697908388501504/3q^4 +
13412366082401583104/3q^3 - 103692674008647256832/15q^2$\\$ +
9998443905803681024/3q$ \\ 
$c_{16}^{(10)}=2934352776960q^{16} -
41801425881600q^{15} + 247247136328192q^{14} - 779579603325952q^{13}$\\$ +
1447311237172736q^{12} - 1455008875567104q^{11} -
386203604227584q^{10} - 11585115503070208/3q^9$\\$ + 5851443127592192q^8 +
3711330462883151104/15q^7 - 1646276274944524544/15q^6 $ \\ $  -
7038770289879316480q^5 + 72320179932609938176/3q^4 -
193773243490858548224/15q^3$\\$ - 251833309601144910592/5q^2 +
60569364218487037184q$ \\ 
$c_{17}^{(10)}=5207010082560q^{17} -
79484527964160q^{16} + 508664327018496q^{15} - 1764964370434048q^{14}
$\\$+ 3679402474964736q^{13} - 4372487916998656q^{12} +
792697521304832q^{11} - 9363099001265152q^{10}$\\$ -
43826578490797184/3q^9 + 8577550308943181056/15q^8 +
28219659363435557888/15q^7$\\$ - 375358120756778684416/15q^6 +
42682963471399987328/3q^5 + 1923822031694136063488/5q^4$\\$ -
20539071570523947907328/15q^3 + 8971213908296949751552/5q^2 -
799998504590970669312q$ \\ 
$c_{18}^{(10)}=8998464034560q^{18} -
146528416869120q^{17} + 1009106692323840q^{16} -
3822882507487232q^{15}$\\$ + 8848871709642496q^{14} -
36520718842539776/3q^{13} + 18743142681086720/3q^{12}$\\$ -
62988187629452800/3q^{11} - 81387924415011456q^{10} +
15571764869132276608/15q^9$\\$ + 383120271984801356288/45q^8 -
234112884730147594496/5q^7 - 13524450156382699305344/45q^6 $\\$+
12823863053571105585024/5q^5 - 249608017584109802434688/45q^4$\\$  -
23073655006552418847872/15q^3 + 893671518686969433181184/45q^2 -
56709341766334119652096/3q$ \\ 
$c_{19}^{(10)}=15180194789760q^{19} -
262646253492480q^{18} + 1937297924305920q^{17} -
7959422215602176q^{16}$\\$ + 20268674276510464q^{15} -
95184859339027456/3q^{14} + 74003803994501888/3q^{13}$\\$ -
139719129016333568/3q^{12} - 242458510757969792q^{11} +
4426762990994442496/3q^{10}$\\$ + 23718695776047253120q^9 -
375197918700176317696/15q^8 - 7237483465040656385536/5q^7$\\$ +
85047934197066602226944/15q^6 + 17815613791796507938176q^5 -
2607613209957224783747584/15q^4 $\\$+ 7100994379431891533783168/15q^3 -
8373915013831095585411584/15q^2$\\$ + 236513875791114976989696q$ \\
$c_{20}^{(10)}=25050011226240q^{20} - 458902348544640q^{19} +
3610245489776640q^{18} - 15993366595831680q^{17}$\\$ +
44460160450820608q^{16} - 78084468994640896q^{15} +
78194195639666944q^{14}$\\$ - 320963319731515136/3q^{13} -
570503588763901312q^{12} + 21341525437006926976/15q^{11} $\\$+
51382154702027327744q^{10} + 757401336790171922432/5q^9 -
33169991145745310954752/9q^8 $\\$- 11264809722306476971136/3q^7 +
8724305421769817690595712/45q^6$\\$ - 13644519855475631409950336/15q^5 +
11216450343505848520675328/9q^4$\\$ + 11162831891410885080944768/5q^3 -
126554341938026732203131904/15q^2$\\$ + 6970666668081782114190720q$ \\
\label{tab10} 
\end{longtable}
\end{center}

\subsection{The  HT expansion of $\chi_{12}(v;d)$}

\begin{center}
\begin{longtable}{l}
\caption{ The coefficients $c^{(12)}_n(d)$ of the HT series expansion in
 powers of $v={\rm tanh}K$ for the  susceptibility
 $\chi_{12}(v;d)=\sum_{n=0,\infty} c^{(12)}_n(d) v^n $ of the 
 spin-1/2 Ising model on a (hyper)-simple-cubical lattice of 
 dimensionality $d$ as polynomials in $q = 2d$.}
 \\
\endhead
\hline \multicolumn{1}{|r|}{{Continued on next page}} \\ 
\hline
\endfoot
\hline
\endlastfoot
\hline
 $c_0^{(12)}=-353792$\\ $c_1^{(12)}=-11184128q$ \\
$c_2^{(12)}=-154918400q^2 + 83051264q$ \\ 
$c_3^{(12)}=-1351633920q^3 +
1745895424q^2 - 585693184q$ \\ 
$c_4^{(12)}=-8658773760q^4 +
18669181440q^3 - 12553458432q^2 + 1241679616q$ \\
$c_5^{(12)}=-44288864256q^5 + 136435353600q^4 - 140287833088q^3 +
28301658112q^2 + 19680018432q$ \\ 
$c_6^{(12)}=-190551588864q^6 +
769929538560q^5 - 1082384808960q^4 + 386087227392q^3 $\\$+ 264833080832q^2
  - 88499933952q$ \\ 
$c_7^{(12)}=-714269736960q^7 + 3587327557632q^6 -
6469828485120q^5 + 3612002744320q^4 $\\$  + 1473372426240q^3 -
380443199488q^2 - 1279808036864q$ \\ 
$c_8^{(12)}=-2391664404480q^8 +
14393298296832q^7 - 31931620611072q^6 + 25342290355200q^5 $\\$  +
2623725271552q^4 + 1090349782016q^3 - 14344234480128q^2 -
840662029824q$ \\ 
$c_9^{(12)}=-7287432355840q^9 + 51178341826560q^8 -
135515864580096q^7 + 142605665599488q^6 $\\$  - 25221009321984q^5 +
49670513813504/3q^4 - 88730929958912/3q^3 - 616452990285824/3q^2 $\\$  +
626694374567936/3q$ \\ 
$c_{10}^{(12)}=-20494654415872q^{10} +
164641056652800q^9 - 508612988841984q^8 + 674985995314176q^7 $\\$  -
292637469406464q^6 + 319963052438272/3q^5 + 933634476506368/3q^4 -
6908597542308352/3q^3 $\\$  + 5810927082614528/3q^2 + 712547619336960q$ \\
$c_{11}^{(12)}=-53792941017088q^{11} + 486701276938240q^{10} -
1723075981086720q^9 + 2779814633496576q^8 $\\$  - 1875623051387904q^7 +
627651365216256q^6 + 3038947079074816q^5 - 36343321694500864/3q^4 $\\$  -
4541280199228416q^3 + 136094192707704832/3q^2 - 33234155640999936q$ \\
$c_{12}^{(12)}=-132951019568640q^{12} + 1338069948959744q^{11} -
5351547829921280q^{10} + 10209312290311680q^9 $\\$  - 9262955488999680q^8 +
3645106505936896q^7 + 15429321176359936q^6 - 118905258348704768/3q^5 $\\$  -
121040226141923840q^4 + 1218820247041000448/3q^3 -
150449878355645952q^2 $\\$  - 234798668068850432q$ \\
$c_{13}^{(12)}=-311662549094400q^{13} + 3454012141891584q^{12} -
15422938056873984q^{11} $\\$  + 34059565585510400q^{10} -
38684966301057024q^9 + 19465049565298688q^8 + 56854453518541824q^7 $\\$  -
85703458485491712q^6 - 2218501762470732800/3q^5 +
3619778929353328640/3q^4  $\\$ + 12274347069953881088/3q^3 -
31957947017706407936/3q^2 + 6191270976593217536q$ \\
$c_{14}^{(12)}=-697110544281600q^{14} + 8436084083481600q^{13} -
41644591004936192q^{12} $\\$  + 104685760328953856q^{11} -
142679933981325312q^{10} + 91914463877743104q^9 +
164968308753550848q^8 $\\$  - 97434386494613504q^7 -
8225702434343863552/3q^6 - 5025700933230147328/3q^5 $\\$  +
129303595089884297984/3q^4 - 222033723027268106240/3q^3 -
28411816242815580928/3q^2 $\\$  + 68608468716171185408q$ \\
$c_{15}^{(12)}=-1495254499614720q^{15} + 19618972950528000q^{14} -
106183823149867008q^{13} $\\$  + 299756354419372032q^{12} -
476680111749890048q^{11} + 383612063412850688q^{10}  $\\$ +
373962329685188608q^9 + 133439709965193216q^8 -
21730914258414602240/3q^7  $\\$ - 431930711865523726336/15q^6 +
559493978162692917248/3q^5 + 163698341420861728768/3q^4  $\\$ -
4707218563575137841152/3q^3 + 41124375194403225552896/15q^2 -
4136125039092031854592/3q$ \\ 
$c_{16}^{(12)}=-3088599584878080q^{16} +
43673418631065600q^{15} - 257320550871659520q^{14} $\\$  +
806768308656407552q^{13} - 1467088196610935808q^{12} +
1433944886486448128q^{11} $\\$  + 556492490967948288q^{10} +
1183366786353330176q^9 - 41636447435305676288/3q^8 $\\$  -
415567921111926978560/3q^7 + 2011839579073394639872/5q^6 +
7952031547616848432640/3q^5 $\\$  - 37897912866498410360320/3q^4 +
13077014203165422923776q^3 + 61638346384830787652608/5q^2 $\\$  -
21409488905851651325440q$ \\ 
$c_{17}^{(12)}=-6166074284697600q^{17} +
93474073271255040q^{16} - 595872544132915200q^{15} $\\$  +
2055801777383424000q^{14} - 4210415450621685760q^{13} +
4873621078494318592q^{12} $\\$  - 220414900643438592q^{11} +
4510103641488453632q^{10} - 15880150474798704640q^9 $\\$  -
450737986045982908416q^8 + 340344532355091195904/15q^7 +
221729576541278514829312/15q^6 $\\$  - 32909314424528161202176q^5 -
351345139137285184575488/3q^4 + 8625784092226962208342016/15q^3 $\\$  -
4070271793983798697129984/5q^2 + 374097575631435887130624q$ \\
$c_{18}^{(12)}=-11934548304806400q^{18} + 193080915347523840q^{17} -
1324578901007831040q^{16} $\\$  + 4989686958615091200q^{15} -
11372156119571563008q^{14} + 45791527096812011008/3q^{13} $\\$  -
17458751461905714688/3q^{12} + 41937490887739925504/3q^{11} +
31933338431806389760/3q^{10} $\\$  - 1153805055301468792576q^9 -
30373341839243953702400/9q^8 + 692581858601325041179904/15q^7  $\\$ +
2590377436954334382597376/45q^6 - 3932407510656106494470912/3q^5 $\\$  +
33593400710271005522988032/9q^4 - 28199426216958322509515264/15q^3 $\\$  -
293998710704524244730689536/45q^2 + 22806424112480971221711104/3q$ \\
$c_{19}^{(12)}=-22455553019827200q^{19} + 386163946273996800q^{18} -
2837525699808184320q^{17} $\\$  + 11593375867440783360q^{16} -
29118875151207942144q^{15} + 133633187409078124544/3q^{14} $\\$  -
87651101589051000832/3q^{13} + 122420035837644673024/3q^{12} $\\$  +
378885862990807923712/3q^{11} - 2460433180576227344384q^{10} -
46949320264420114518016/3q^9  $\\$ + 272796345569841313445888/3q^8 +
11043097785769283749154816/15q^7 $\\$  - 76587116791527836322676736/15q^6 +
535138423569783282610176q^5 $\\$  + 71151625301216987466575872q^4 -
1141571343228168064195402752/5q^3 $\\$ 
 + 4262414727355334828066523136/15q^2 - 123164411164904424887457792q$ \\
$c_{20}^{(12)}=-41170315007028480q^{20} + 749910487155648000q^{19} -
5877540461959246080q^{18} $\\$  + 25896572589293579520q^{17} -
71102932389431354880q^{16} + 122229074249888176128q^{15} $\\$  -
108969528004928226304q^{14} + 357322559527178232832/3q^{13} +
458200402892033175296q^{12} $\\$  - 66914070629597098638848/15q^{11} -
47955624592773370635520q^{10} + 80984110082112633531648q^9 $\\$  +
134754839963129354103603712/45q^8 - 118043508845640078600391168/15q^7 $\\$ 
- 3966222753465191519663303168/45q^6 +
1733809313650138643157406208/3q^5 $\\$  -
52123638124221675719579086592/45q^4 - 673442963130624721555058688/5q^3 $\\$ 
+ 49780845770293016030065316096/15q^2 - 3131394174087958521457421568q$
\\ 
 \label{tab12} 
\end{longtable}
\end{center}

\subsection{The  HT expansion of $\chi_{14}(v;d)$}

\begin{center}
\begin{longtable}{l}
\caption{ The coefficients $c^{(14)}_n(d)$ of the HT series expansion in
 powers of $v={\rm tanh}K$ for the  susceptibility
 $\chi_{14}(v;d)=\sum_{n=0,\infty} c^{(14)}_n(d) v^n $ of the 
 spin-1/2 Ising model on a (hyper)-simple-cubical lattice of 
 dimensionality $d$ as polynomials in $q = 2d$.}
 \\
\endhead
\hline \multicolumn{1}{|r|}{{Continued on next page}} \\ 
\hline
\endfoot
\hline 
\endlastfoot
\hline
 $c_0^{(14)}=22368256$\\ 
$c_1^{(14)}=951878656q$ \\
$c_2^{(14)}=17171485696q^2 - 9061682176q$ \\
$c_3^{(14)}=190346960896q^3 - 242153541632q^2 + 81035140096q$ \\
$c_4^{(14)}=1519012888576q^4 - 3228372738048q^3 + 2173094684672q^2 -
241910319104q$ \\ 
$c_5^{(14)}=9522179620864q^5 - 28942242217984q^4 +
29810692870144q^3 - 6697031569408q^2 $\\$  - 3638385785856q$ \\
$c_6^{(14)}=49521284890624q^6 - 197607542865920q^5 +
278159316692992q^4 - 106288041009152q^3 $\\$  - 61351205630976q^2 +
25207593720832q$ \\ 
$c_7^{(14)}=221721122013184q^7 -
1100726949249024q^6 + 1986223294693376q^5 - 1154500961435648q^4 $\\$  -
426102323097600q^3 + 245451202273280q^2 + 271935058287616q$ \\
$c_8^{(14)}=877590034038784q^8 - 5224913761435648q^7 +
11588424057585664q^6 - 9410419127386112q^5 $\\$  - 1000744824053760q^4 +
1075295221833728q^3 + 4029547520798720q^2 - 338914506043392q$ \\
$c_9^{(14)}=3132002019784704q^9 - 21776871451738112q^8 +
57607925238919168q^7 - 61388804936712192q^6 $\\$  + 9491513408580608q^5 +
6447340566738944/3q^4 + 44754385667900416/3q^3 +
154173876921395200/3q^2 $\\$  - 182161361411537920/3q$ \\
$c_{10}^{(14)}=10232670234040320q^{10} - 81443029784979456q^9 +
251210594330341376q^8 $\\$  - 335497615196946432q^7 + 135461448552674304q^6
- 35792119632002048/3q^5 - 222702217883454464/3q^4 $\\$  +
2362461652009822208/3q^3 - 2456278422725023744/3q^2 -
111726801462610944q$ \\ 
$c_{11}^{(14)}=30973029286748160q^{11} -
277822048171008000q^{10} + 981629865368946688q^9 $\\$  -
1587563073560690688q^8 + 1020762474263646208q^7 -
209702033422557184q^6 - 1173173825248046080q^5 $\\$  +
16683124950400256000/3q^4 - 1176998384765688832q^3 -
43334277608785086464/3q^2 $\\$  + 11767439556985869312q$ \\
$c_{12}^{(14)}=87695613221928960q^{12} - 875516256400220160q^{11} +
3493448597460332544q^{10} $\\$  - 6664923981790646272q^9 +
5837821311797616640q^8 - 5668522943666667520/3q^7 $\\$  -
22994941740443248640/3q^6 + 25365453641949280256q^5 +
103451675164792563712/3q^4 $\\$  - 526401211985641717760/3q^3 +
312941970439251582976/3q^2 + 65165416306793678848q$ \\
$c_{13}^{(14)}=234088827521064960q^{13} - 2574836787374161920q^{12} +
11467569791287357440q^{11} $\\$  - 25287282865045250048q^{10} +
27956165485024573440q^9 - 38557986919775731712/3q^8 $\\$  -
104508826006886752256/3q^7 + 255375882161599811584/3q^6 +
952361429739659500544/3q^5 $\\$  - 846995991975855159296q^4 -
1139738243810908156928q^3 + 12549696909467484780544/3q^2 $\\$  -
2597769507194729141248q$ \\ 
$c_{14}^{(14)}=592949907428597760q^{14} -
7125199541722030080q^{13} + 35076261083993886720q^{12} $\\$  -
87958340403507830784q^{11} + 117329974060356933632q^{10} -
71843845823446751232q^9 $\\$  - 122586910874966786048q^8 +
669004110698680303616/3q^7 + 4744396665028853771264/3q^6 $\\$  -
3669495689471793658880/3q^5 - 57018764157647422594048/3q^4 +
41105792891731103113216q^3 $\\$  - 25186090822497939331072/3q^2 -
24319035703272885756928q$ \\ 
$c_{15}^{(14)}=1433048994515435520q^{15} -
18678993309190717440q^{14} + 100804343240191549440q^{13} $\\$  -
283697078951770030080q^{12} + 443223271221351739392q^{11} -
1033637533951132155904/3q^{10} $\\$  - 1011217207474809448448/3q^9 +
460483675184064487424q^8 + 5565906744552960131072q^7 $\\$  +
43919638020575731011584/5q^6 - 347024996061061408676864/3q^5 +
239690733943793419577344/3q^4 $\\$  + 1866879933655208536607744/3q^3 -
6335397272764050983266304/5q^2 $\\$  + 2000581556284493794393088/3q$ \\
$c_{16}^{(14)}=3319867725632901120q^{16} - 46653317999297740800q^{15} +
274061753900784107520q^{14} $\\$  - 856289832024370053120q^{13} +
1533976830053824094208q^{12} - 4377264783509685641216/3q^{11} $\\$  -
1930294775907870310400/3q^{10} + 2043781452722642796544/3q^9 +
44749069561807159070720/3q^8 $\\$  + 368688640717095980146688/5q^7 -
6020966558926158905196544/15q^6 - 959320791731322801360896q^5 $\\$  +
21141616501494931472162816/3q^4 - 144951016231397003748696064/15q^3 $\\$  -
11761559674844284219748352/5q^2 + 8971329699132498653917184q$ \\
$c_{17}^{(14)}=7401483245876689920q^{17} - 111549432110458552320q^{16} +
708955002035236392960q^{15} $\\$  - 2436938752035708026880q^{14} +
4926616716589229905920q^{13} - 5571945225556920684544q^{12} $\\$  -
164482664925404280832q^{11} + 542451596580084125696/3q^{10} +
90500305739342358821888/3q^9 $\\$  + 4935275039197882432403456/15q^8 -
3828311185429464106921984/5q^7 $\\$  - 43083052926904186971451392/5q^6 +
91947713254873005894375424/3q^5 $\\$  + 150183300690139346326908928/5q^4 -
1386080926679917894740688896/5q^3 $\\$  + 2150316704653739855171315712/5q^2
- 204220289203880776017288192q$ \\
$c_{18}^{(14)}=15934593393411025920q^{18} - 256381800437179069440q^{17}
+ 1753506681434233835520q^{16} $\\$  - 6580315733798396805120q^{15} +
14826378622802061465600q^{14} - 19494270830426973841408q^{13} $\\$  +
5840961809226396641280q^{12} - 12379160817027722752000/3q^{11} +
114757699742723572333568/3q^{10} $\\$  + 16488703030550456426288128/15q^9 +
9566178324343212097599488/45q^8 $\\$  - 191473112196390312892758016/5q^7 +
1539992983726049687744905216/45q^6 $\\$  +
10601363112007581786085537792/15q^5 -
114323904676714821742061170688/45q^4 $\\$  +
34146144587643522501476139008/15q^3 +
103223347344087124949760379904/45q^2 $\\$  -
10901839332941488863023727616/3q$ \\
$c_{19}^{(14)}=33226125696362987520q^{19} - 568427635729473269760q^{18}
+ 4164157054396259312640q^{17} $\\$  - 16947979998253087703040q^{16} +
42132111162172651745280q^{15} - 63257522133446069288960q^{14} $\\$  +
36355926368650676676608q^{13} - 74447257194277928488960/3q^{12} -
54484321215182590646272/3q^{11} $\\$  + 9097409010673707965517824/3q^{10} +
24262474856816972871058432/3q^9 $\\$  - 1714006446732822316017821696/15q^8 -
4665520708535731448468758528/15q^7 $\\$  +
61022631460493300164426153984/15q^6 - 5972682674402348291090053120q^5 $\\$ 
- 468681727160463278326158270464/15q^4 +
1905595883401702552479227158528/15q^3 $\\$  -
2524382647775783152293231110144/15q^2 +
74776398708463644831068077056q$ \\
$c_{20}^{(14)}=67276931969120501760q^{20} - 1219446678396708372480q^{19}
+ 9528887956728128378880q^{18} $\\$  - 41822293160168382044160q^{17} +
113760507839982661263360q^{16} - 192236223612750884290560q^{15} $\\$  +
156489073462191969669120q^{14} - 110241775369845605625856q^{13} -
919955615111802494734336/3q^{12} $\\$  + 36119687323866478101979136/5q^{11}
+ 38236807443309719843198976q^{10} $\\$  - 1191331056719902439094376448/5q^9
- 19547313607379395121578246144/9q^8 $\\$  +
180510166791180821800942637056/15q^7 +
1605759665205848894990726950912/45q^6 $\\$  -
5794973282407102465659826425856/15q^5 +
8663267176107152845551619469312/9q^4 $\\$  -
2296147211477278179472071303168/5q^3 -
22163092582660818160691985946624/15q^2 $\\$  +
1673593062235423907376589068288q$ \\
 \label{tab14} 
\end{longtable}
\end{center}

\subsection{The  HT expansion of $\chi_{16}(v;d)$}

\begin{center}
\begin{longtable}{l}
\caption{ The coefficients $c^{(16)}_n(d)$ of the HT series expansion in
 powers of $v={\rm tanh}K$ for the  susceptibility
 $\chi_{16}(v;d)=\sum_{n=0,\infty} c^{(16)}_n(d) v^n $ of the 
 spin-1/2 Ising model on a (hyper)-simple-cubical lattice of 
 dimensionality $d$ as polynomials in $q = 2d$.}
\\
\endhead
\hline \multicolumn{1}{|r|}{{Continued on next page}} \\ 
\hline
\endfoot
\hline
\endlastfoot
\hline
 $c_0^{(16)}=-1903757312$\\ 
$c_1^{(16)}=-104932671488q$ \\
$c_2^{(16)}=-2389096202240q^2 + 1247014436864q$ \\
$c_3^{(16)}=-32769353973760q^3 + 41234501730304q^2 - 13779811467264q$
\\ $c_4^{(16)}=-318434742599680q^4 + 669638839173120q^3 -
451194468442112q^2 + 54187912339456q$ \\
$c_5^{(16)}=-2398328708005888q^5 + 7216293157273600q^4 -
7443684913840128q^3 $\\$  + 1796451694804992q^2 + 810471469645824q$ \\
$c_6^{(16)}=-14814711540744192q^6 + 58553079162798080q^5 -
82516960690176000q^4 $\\$  + 33131931010138112q^3 + 16568564934164480q^2 -
7826521298747392q$ \\ 
$c_7^{(16)}=-78001193991536640q^7 +
383758520681496576q^6 - 692893232058204160q^5 $\\$  + 415590082033745920q^4
+ 138050755428352000q^3 - 112563191669850112q^2 $\\$  - 67250352764125184q$
\\ $c_8^{(16)}=-359881375772160000q^8 + 2124537375912247296q^7 -
4712048562925756416q^6 $\\$  + 3901175774945075200q^5 +
382668819973588992q^4 - 830697243533570048q^3 $\\$  - 1253425152787441664q^2$
$+ 246687187735013376q$ \\ 
$c_9^{(16)}=-1485475101516595200q^9 +
10246573574686310400q^8 - 27091221120030670848q^7 $\\$  +
29207910483003244544q^6 - 4295104640277151744q^5 -
12340496168646148096/3q^4 $\\$  - 19033354976936984576/3q^3 -
44592692556107743232/3q^2 + 62059824450177499136/3q$ \\
$c_{10}^{(16)}=-5573914967367352320q^{10} + 44032574907530035200q^9 -
135681242051050831872q^8 $\\$  + 182431884556188581888q^7 -
70873489496818745344q^6 - 32360324729745539072/3q^5  $\\$ +
67706322643199107072/3q^4 - 919115973967527559168/3q^3 +
1110469663689548767232/3q^2 $\\$  + 4028078481045454848q$ \\
$c_{11}^{(16)}=-19254430629988270080q^{11} + 171497359126619750400q^{10}
- 605111828654049361920q^9  $\\$ + 982109969155263234048q^8 -
613201437461336227840q^7 + 52147748159124078592q^6 $\\$  +
540787451457111064576q^5 - 8293323481894075432960/3q^4 +
1601119273086217224192q^3 $\\$  + 15799373824615516340224/3q^2 -
4836104293955961716736q$ \\ 
$c_{12}^{(16)}=-61858597238822092800q^{12} +
613495164956018442240q^{11} - 2443811528379954585600q^{10} $\\$  +
4669086468011990999040q^9 - 3997355520769368145920q^8 +
3048460248858627997696/3q^7 $\\$  + 13023044456997156921344/3q^6 -
48507816777292733022208/3q^5 - 24805283810470472028160/3q^4  $\\$ +
83595989933429336940544q^3 - 191790280034196466143232/3q^2 -
19678447759486369759232q$ \\ 
$c_{13}^{(16)}=-186378077798827622400q^{13}
+ 2037305764689922621440q^{12} - 9056133528765221437440q^{11}  $\\$ +
19970129771694656716800q^{10} - 21679251241263612198912q^9 +
26990843334412000231424/3q^8 $\\$  + 70554377277375853297664/3q^7 -
213211678483330007105536/3q^6 - 139778987713044081672192q^5 $\\$  +
1664707021142943756255232/3q^4 + 839278592660028565356544/3q^3$\\$ -
5660286687163968965574656/3q^2   + 1258500333536581549654016q$ \\
$c_{14}^{(16)}=-530309836482532147200q^{14} +
6335138766411898060800q^{13} - 31121484056466262917120q^{12} $\\$  +
77967410300989969858560q^{11} - 102461246405166576254976q^{10} +
59417543428498750226432q^9 $\\$  + 97390176985860741332992q^8 -
762964738187005595942912/3q^7 - 2771862731613706912448512/3q^6 $\\$  +
1762940830895704098471936q^5 + 27268466707857705964199936/3q^4 $\\$  -
73144748403820602006495232/3q^3+ 29494003189156796786991104/3q^2 $\\$  +
9859878527425768111505408q$ \\
$c_{15}^{(16)}=-1433313857854681251840q^{15} +
18579093268881997824000q^{14} - 100041765469328009134080q^{13} $\\$  +
281099599114713305579520q^{12} - 433684248549726146396160q^{11} +
325666378885024346406912q^{10} $\\$  + 314189928509364533886976q^9 -
778640166888945468899328q^8 - 12345110484432923427930112/3q^7 $\\$  -
6029605745496073616162816/15q^6 + 74183532794057737351725056q^5 $\\$  -
299336476697102553835700224/3q^4 - 812355558976921220939382784/3q^3 $\\$  +
10079179148554636101429231616/15q^2 - 1113855886758632875032739840/3q$ \\
 $c_{16}^{(16)}=-3698167731092948367360q^{16} +
51697759510997113651200q^{15} - 302989963080538533273600q^{14} $\\$  +
944731356246521622405120q^{13} - 1674213903991364760268800q^{12} +
1552712554992677806497792q^{11} $\\$  + 720447356096041891672064q^{10} -
6282927250512377689477120/3q^9 - 13916774158254107683837952q^8  $\\$ -
109261731058784289328144384/3q^7 + 5247550300007197475660517376/15q^6 $\\$ 
+ 255075570728706442667266048q^5 - 12816733701499485987675908096/3q^4 $\\$ 
+ 21265133695803460802539581440/3q^3 -
2749516121165670024115259392/5q^2 $\\$  - 4346747332669088027247523840q$ \\
$c_{17}^{(16)}=-9147798671639519232000q^{17} +
137186452951627556782080q^{16} - 869813208451446113894400q^{15} $\\$  +
2982818202880574005248000q^{14} - 5973178395596154214809600q^{13} +
6619544164878738447728640q^{12} $\\$  + 481000794676112127098880q^{11} -
4899603407702683328774144q^{10} - 110565361480733286376570880/3q^9  $\\$ -
704983319085229368702533632/3q^8 + 16103452583665361645089193984/15q^7 $\\$ 
+ 75829256330761417932755566592/15q^6 -
77977148669047737921882554368/3q^5 $\\$  + 1811968389847140769282719744q^4 +
2257670174798245419856825286656/15q^3 $\\$  -
1302173999037493066633818865664/5q^2 +
128088962284953555125162901504q$ \\
$c_{18}^{(16)}=-21773726436257454489600q^{18} +
348689272216526236385280q^{17} - 2379062154498192511303680q^{16} $\\$  +
8904943167941543976960000q^{15} - 19894429006009620467220480q^{14} +
25707317955109416860549120q^{13} $\\$  - 6350952062228959616368640q^{12} -
8916914984431825074126848q^{11} - 219363621455006506598907904/3q^{10}
 $\\$ - 3025488708891444056995053568/3q^9 +
16821728067680996475046395904/9q^8  $\\$ +
464207279270231673467090960384/15q^7 -
3281882880491767934129981210624/45q^6 $\\$  -
1193128885900196613472098156544/3q^5 +
16617460149856001792874156007424/9q^4 $\\$  -
32155737405107925626486996885504/15q^3 -
32556846223405514512001694785536/45q^2 $\\$  +
6046576182698989570541898727424/3q$ \\
$c_{19}^{(16)}=-50030037863288065228800q^{19} +
852101290521418412851200q^{18} - 6227047101542461421322240q^{17} $\\$  +
25275614189702156621905920q^{16} - 62352271739853593258557440q^{15} +
92187604552840911877242880q^{14} $\\$  - 47835588891081063187742720q^{13} -
4548900794238086113460224q^{12} $\\$  - 237947979009622578034180096/3q^{11} 
- 10366035347757135714664644608/3q^{10} $\\$ -
1265874199130683595892228096q^9   + 123225127752271144007575142400q^8 $\\$ +
634660532275614856202776477696/15q^7   -
48783751300574925991281498914816/15q^6  $\\$+
23362355833110189429704435826688/3q^5   +
40370908413300708220306030919680/3q^4 $\\$-
1202743242344609002546580936851456/15q^3   +
1712098811966697109750802892783616/15q^2$\\$ -
52084653311579531984812997640192q$ \\
$c_{20}^{(16)}=-111284638050657492664320q^{20} +
2008604722638854258688000q^{19} $\\$  - 15657114688558678873128960q^{18} +
68528531116644896241008640q^{17}  $\\$ - 185091246922911993627525120q^{16} +
308456737644166806503424000q^{15} $\\$  - 233888851657298026447503360q^{14}
+ 67391655489588897810415616q^{13} $\\$ 
 + 141274859918430596363763712q^{12}
- 50591878074073047993295863808/5q^{11} $\\$  -
24823899558370202782000799744q^{10} +
362123860232776118591214739456q^9 $\\$  +
64086239757063227627779602399232/45q^8 -
207863878816724357223254796271616/15q^7 $\\$  -
283156467508485410552545878499328/45q^6 +
828660290657745177421922636824576/3q^5  $\\$ -
37071865469327969561767228813586432/45q^4 +
660261762816077560188055602307072q^3  $\\$              +
10536441619319090528029264686166016/15q^2 -
1039099543218814255033842395971584q$ \\ 
\label{tab16} 
\end{longtable}
\end{center}

\end{document}